\documentclass[11pt,twoside]{article}
\usepackage{asp2010}
\resetcounters

\begin{document}

\title{On The Formation of Double White Dwarfs: Reevaluating How We Parametrize the Common Envelope Phase}
\author{T.E. Woods$^1$, N. Ivanova $^1$, M. van der Sluys$^2$, and S. Chaichenets$^3$
\affil{$^1$University of Alberta, Dept.\ of Physics, 11322-89 Ave, Edmonton, AB, T6G 2E7, Canada}
\affil{$^2$Radboud University Nijmegen, P.O.\ Box 9010, NL-6500 GL Nijmegen, The Netherlands}
\affil{$^3$University of Alberta, Dept.\ of Mathematical and Statistical Sciences, CAB, Edmonton, AB, T6G 2G1, Canada}}

\begin{abstract}
 One class of compact binaries of special interest is that of double white dwarfs (DWDs). For many of these systems, the exact nature
of the evolutionary channels by which they form remains uncertain. The canonical explanation calls
for the progenitor binary system to undergo two subsequent mass-transfer events, both of which are
unstable and lead to a common envelope (CE) phase. However, it has been shown that if both CE events
obey the standard $\alpha _{\mathrm{CE}}$ prescription, it is not possible to reproduce all of the observed systems. As an
alternative prescription, the $\gamma$-formalism was proposed, which parametrizes the fraction of angular
momentum carried away with mass loss, in contrast to the $\alpha _{\mathrm{CE}}$ prescription, which parametrizes energy loss. We demonstrate that the
$\gamma$-prescription is also inadequate in describing the evolution of an arbitrary DWD binary; clearly we require a deeper understanding of the physical mechanisms underlying their formation. We then present a detailed model for the evolution of Red Giant -- Main Sequence binaries during the first episode of mass transfer, and demonstrate that their evolution into DWDs need not arise through two phases of dynamical mass loss.  Instead, the first episode of dramatic mass loss may be stable, non-conservative mass transfer. The second phase is then well described by the $\alpha _{\mathrm{CE}}$ prescription. We find that the considered progenitors can reproduce the properties of the observed helium DWDs in which the younger component is the more massive.

\end{abstract}

\section{Problems with Parametrizing CE Evolution}

In order to explain the dramatic loss of mass and angular momentum needed to form the majority of observed close binaries with at least one degenerate component, \cite{Pac76} and \cite{Ost76} first suggested such systems undergo a common envelope (CE) phase during their evolution. In this event one component expands on the giant branch to engulf the other leading to spiral-in and the removal of the envelope. In the standard picture, the consequent orbital shrinkage is parametrized in a simple manner in terms of the mass lost and an efficiency factor $\alpha_{\mathrm{CE}}$ (though see Ivanova, this volume):

\begin{equation}\label{alpha}
\alpha _{\mathrm{CE}} \left( \frac{Gm_{\mathrm{d,c}}m_{\mathrm{a}}}{2a_{\mathrm{f}}} - \frac{Gm_{\mathrm{d}}m_{\mathrm{a}}}{2a_{\mathrm{i}}} \right) = \frac{Gm_{\mathrm{d}}m_{\mathrm{d,e}}}{\lambda r_{\mathrm{d}}}
\end{equation}

\noindent \citep{alpha_ref} where $m_{\mathrm{a}}$ is the mass of the companion, $a_{\mathrm{i}}$ and $a_{\mathrm{f}}$ are the initial and final separation, $r_{d}$ is the radius of the donor at the onset of the event, and $m_{\mathrm{d}}$, $m_{\mathrm{d,c}}$, and $m_{\mathrm{d,e}}$ are the masses of the donor, its core, and its envelope, respectively. 

However, problems soon became apparent in this straight-forward picture, most glaringly in attempting to apply this prescription to the study of double white dwarfs (DWDs). Such binaries are considered one of our best means of constraining CE evolution, since due to their close orbit they must necessarily have passed through a CE as their most recent episode of mass loss. At the same time, their small size (the majority are found as low-mass helium dwarfs) means they could not possibly have formed by any normal process of single-stellar evolution within a Hubble time, implying both components must have come about through substantial binary interaction (i.e. two phases of dramatic mass loss while on the first giant branch). \cite{Nelemans2000} demonstrated that, were such systems to have evolved through two CE phases, many would have needed to expel the initial donor's envelope with a negative efficiency ($\alpha _{\mathrm{CE}}$) in the first event. At the same time, an initial phase of stable, conservative case B mass transfer appeared to be an unlikely scenario \citep{Nelemans2000, Sluys2006}. This at the time suggested that an alternative mechanism may be needed in order to explain such systems. 

In response to the apparent need for a negative efficiency in the first phase of dramatic mass loss, \cite{Nelemans2000} suggested that rather than guessing at the energetics of the event, one could invoke an alternative parametrization in terms of the angular momentum lost:

\begin{equation}\label{gamma}
\frac{\Delta J_{\mathrm{lost}}}{J_{\mathrm{i}}} = 1 - \frac{J_{\mathrm{final}}}{J_{\mathrm{start}}} = \gamma \frac{m_{\mathrm{d,e}}}{m_{\mathrm{d}} + m_{\mathrm{a}}} = \frac{m_{\mathrm{d}} - m_{\mathrm{d,c}}}{m_{\mathrm{d}} + m_{\mathrm{a}}}
\end{equation}

\noindent where the parameter $\gamma$ describes the average specific angular momentum of the binary carried away by the mass shed from the system. This allowed for a significant shrinkage of the orbit in the second episode of mass loss with only a modest reduction or even widening of the orbit in the first phase, dependent on the input masses.  This in turn led \cite{Nelemans2005} to claim that a range $1.5 \lesssim \gamma \lesssim 1.75$ could satisfactorily reproduce {\bf all} post-CE binaries, for specific ranges of initial masses in each case. 

However, while fits for $\gamma$ in this range were found to be common to all those systems used in their calibration, there remains quite a wide range in plausible values of $\gamma$ for each individual system, corresponding to a range of possible progenitor masses. Prior studies \citep[][also see Zorotovic et al, and Davis et al, this volume]{Webbink2008,Zorotovic2010,Davis2010} have demonstrated that in general the use of the $\gamma$ formalism rather poorly constrains the possible set of solutions in reconstructing post-common-envelope binaries. 

The reasoning behind the great multiplicity of solutions offered by such a narrow range of $\gamma$ values becomes clear if we examine the effect of varying $\gamma$ for a fixed system. (Note that the discussion here follows that of \cite{Woods2011arxiv}, $\S$ 3). Consider the parametrization given in equation \ref{gamma}; if we substitute for the initial and final angular momenta:

\begin{equation}
J_{\mathrm{start}} = G^{\frac{1}{2}}\frac{m_{a}m_{d}}{\sqrt{m_{a} + m_{d}}}a_{\mathrm{i}}^{\frac{1}{2}}, \hskip1.0cm J_{\mathrm{final}} = G^{\frac{1}{2}}\frac{m_{\mathrm{d,c}}m_{a}}{\sqrt{m_{\mathrm{d,c}} + m_{a}}}a_{\mathrm{f}}^{\frac{1}{2}}
\end{equation}

\noindent then we can solve for the initial over the final separation:

\begin{equation}\label{afinal}
\left(\frac{a_{\mathrm{f}}}{a_{\mathrm{i}}}\right)^{\frac{1}{2}} = \left(1 - \gamma \frac{1-q_{\mathrm{f}}}{1+q_{\mathrm{1}}}\right)\sqrt{(q_{1}+q_{\mathrm{f}})/(1+q_1)}\frac{1}{q_{\mathrm{f}}}
\end{equation}

\noindent where we use the notation $q_{\mathrm{f}}=m_{\mathrm{d,c}}/m_{d}$ and $q_{1} = m_{a}/m_{d}$. Taking $a_{\mathrm{f}}/a_{\mathrm{i}}$ as a single parameter we can find the derivative of our solution with respect to the value of $\gamma$:

\begin{center}
\begin{equation}
 \frac{\partial a_{\mathrm{f}}/a_{\mathrm{i}}}{\partial \gamma} = -2B(A - \gamma B), \hskip0.5cm A = \sqrt{(q_{1}+q_{\mathrm{f}})/(1+q_1)}\frac{1}{q_{\mathrm{f}}}, \hskip0.5cm B = \frac{1-q_{\mathrm{f}}}{1+q_{\mathrm{1}}}A
\end{equation}
\end{center}

\noindent where we have bundled up all of the mass terms in order to examine the effect of varying $\gamma$, while holding other parameters fixed. For values ($m_{\mathrm{d}}=2.0M_{\odot}$, $m_{\mathrm{d,}} = 0.4M_{\odot}$, $m_{\mathrm{a}} = 1.5 M_{\odot}$) typical of an initial phase of mass loss in the formation of a DWD as prescribed by \cite{Nelemans2000, Nelemans2005}, we see that in the range $1.5 < \gamma < 1.75$ we have $-3.9 < \partial (a_{\mathrm{f}}/a_{\mathrm{i}})/\partial \gamma  < -2.5$. Here a change in $\gamma$ by $\sim 0.25$ leads to $\partial a_{\mathrm{f}}  \sim \partial a_{\mathrm{i}}$. 

\begin{figure}
\begin{center}.
\includegraphics[height=0.3\textheight]{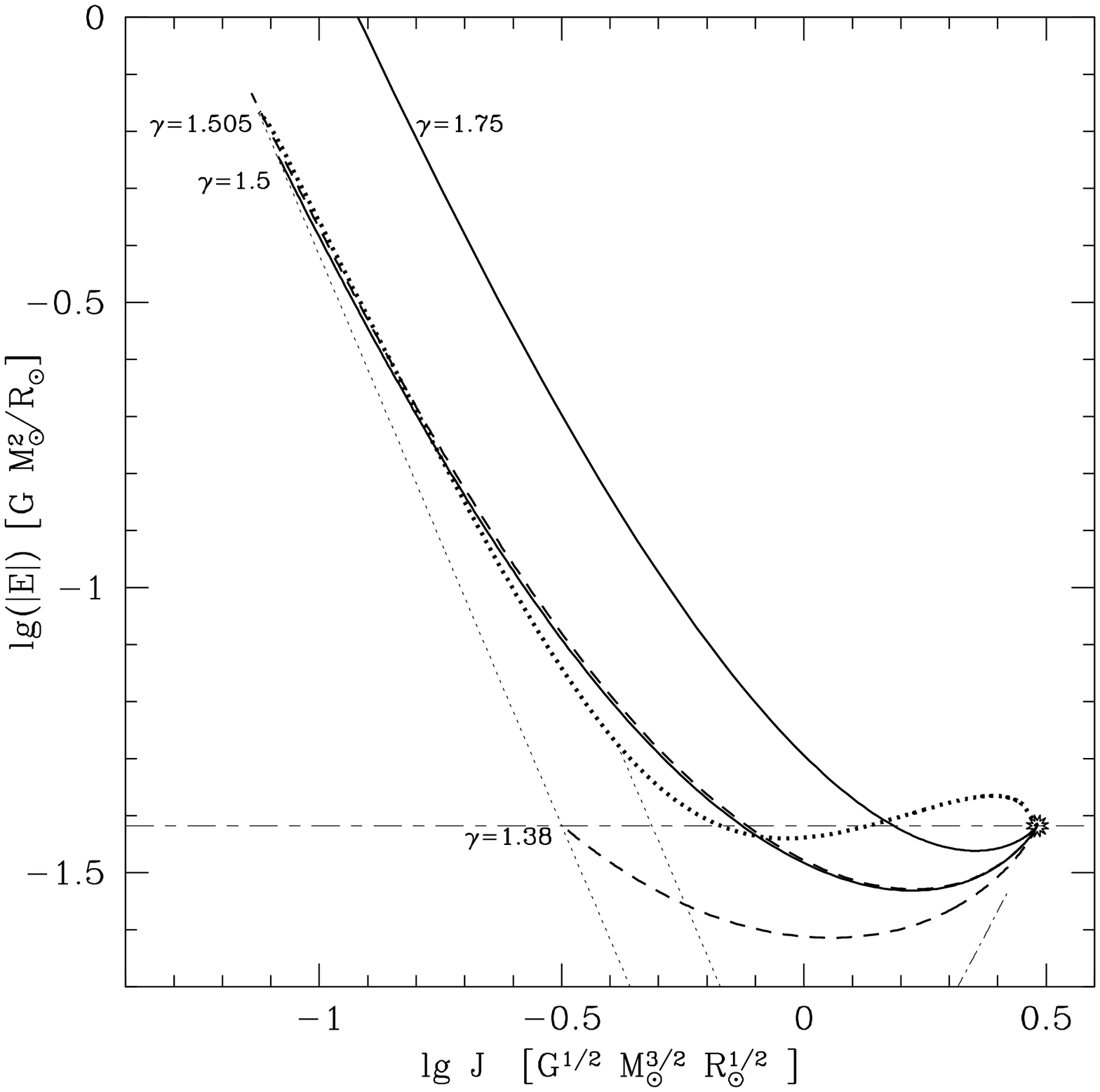} 
\includegraphics[height=0.3\textheight]{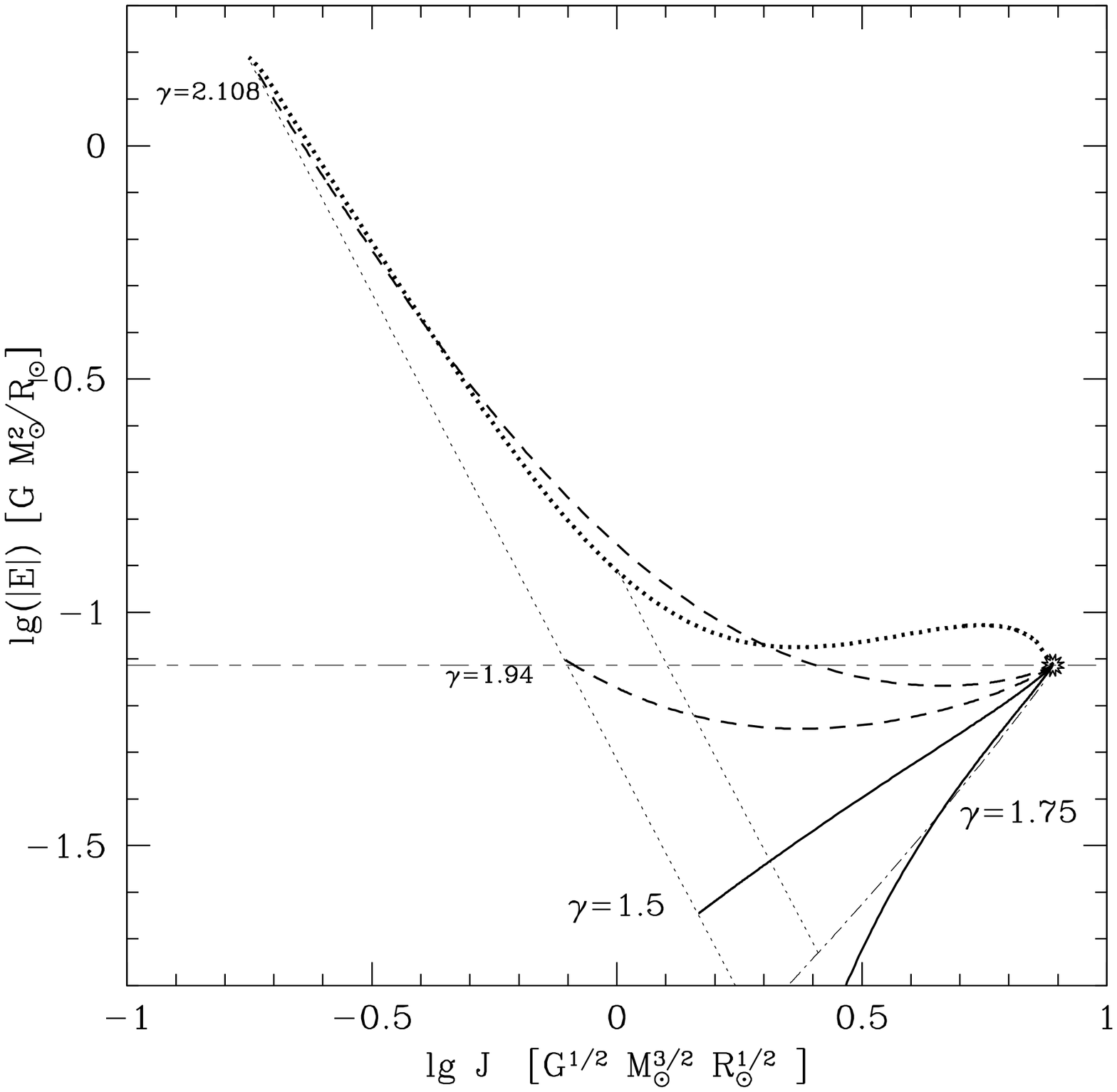}
\caption{Effect of small changes in $\gamma$ on the final period, for $2.0+0.5M_{\odot}$ ({\it left}) and $2.0+1.5M_{\odot}$ binaries ({\it right}). Solid, thin gray lines denote bounds for an energy consistent DWD given by assuming its final orbit must be Keplerian Dash-dotted thin gray line marks the initial energy. For further details see \cite{Woods2011arxiv}.}
\end{center}
\end{figure}

Clearly the final solution is quite sensitively dependent on the value of $\gamma$ chosen, as was pointed out by \cite{Webbink2008}, and in fact an only slightly greater range allows for anything from merger events to greatly expanded orbits. It is then extremely important that the ``correct" value of $\gamma$ be used for any given system. In figure 1, we see in the first panel that for a $2+0.5M_{\odot}$ binary (with a $0.38M_{\odot}$ hydrogen-exhausted core) $\gamma \approx 1.505$ would give a energy-and-angular-momentum consistent DWD system, whereas $\gamma \approx 1.75$ would lead to a merger. If one were evolving this system forward, for example in a population synthesis study, the question becomes which is the appropriate value for the given masses. Similar behaviour can be seen for a system on the verge of an initial episode of mass loss (second panel).

While \cite{Nelemans2005} find $1.5 \lesssim \gamma \lesssim 1.75$ can provide plausible reconstructed initial conditions for all DWD systems, individual binaries can be fit with values of $\gamma$ ranging from 0.5 --$\sim$ 4 for given initial masses, differing from system to system. This relates to the fact that the final separation is very strongly dependent on the envelope mass available prior to the event. However, we have no means of knowing a priori what the true initial masses, and so the correct value of $\gamma$, would be for any known DWD we attempt to reconstruct. The difficulty remains that we have four unknowns: namely the initial separation; the separation after the first mass-loss event which stripped the envelope of the intially more massive component; and the initial masses of both components, while {\it Nature} provides us with only 3 constraints: the final separation, and the final remnant masses. This of course excludes any complicating factors such as the partial accretion by the older degenerate component of wind from its companion; any orbital evolution outside of the two main phases of mass loss (e.g. wind loss); or attempting to reconcile our understanding of the reconstructed evolutionary pathway with the observed difference in ages of the components \citep{Sluys2006}, an uneasy task given our great uncertainty in determining the relative ages of white dwarfs. Without a physical theory underlying the $\gamma$ formalism which assigns a specific value of $\gamma$ for any given set of initial conditions, we cannot invoke angular momentum balance {\it alone} as a constraint upon the evolution of an arbitrary DWD. 

A lingering issue remains the relevant timescale associated with each episode of dramatic mass loss. While the second event must necessarily be a CE phase in order to produce the observed close orbit, the prior evolution is much more uncertain. In particular, the initial phase appears to eject the envelope with only a minor net reduction in the orbit, or even substantial expansion. Although typically presented as a dynamical timescale event, \cite{Sluys2006} as well as \cite{Nelemans2000, Nelemans2005} only restrict themselves to timescales short compared to the nuclear evolution of the donor in their formulation, and in general a ``$\gamma$-prescribed event" could also be thought of as being a much longer, thermal timescale evolutionary phase (Gijs Nelemans, private communication). However, an underlying physical mechanism for some spontaneous ejection of the envelope with little effect on the orbit remains lacking. 

For those DWD binaries in which we see evidence for an initial phase of orbital expansion with mass loss (namely, the older component being less massive), one would expect stable mass transfer (MT) to be a likely candidate. Indeed, though prior studies had claimed to have ruled this out, they did so under the assumption that any potential donor star which had evolved past the base of the giant branch would have too deep a convective envelope to allow for stable mass transfer, leaving a CE the only recourse upon reaching Roche lobe overflow (RLOF). However, we note that MT is increasingly stable with greater core masses \citep{Hjellming1987} and that, particularly if one allows for non-conservative MT, stability is in fact quite possible.
\section{Stable MT followed by a CE}

\subsection{Stability Criteria}

In order for mass transfer to proceed in a stable manner, we require that the response of the donor star to mass loss be such that it remain within its Roche lobe. We can define a response exponent $\zeta$ such that $R \propto M^{\zeta}$, which gives us the requirement $\zeta _{\mathrm{RL}} \leq \zeta _{\mathrm{ad}}$ for stability against dynamical timescale mass loss \citep[where $\zeta _{\mathrm{ad}}$ is the adiabatic response of the donor, see][]{Hjellming1987} and $\zeta _{\mathrm{RL}} \leq \zeta _{\mathrm{th}}$, where $\zeta _{\mathrm{th}}$ is the thermal response of the donor \citep{Soberman1997}. $\zeta _{\mathrm{RL}}$ is of course the response of the Roche lobe to mass loss from the donor; one can easily derive:

\begin{equation}
\zeta _{\mathrm{RL}} = \frac{d\log R_{\mathrm{RL}}}{d \log m_{\mathrm{d}}} = \frac{\partial \ln a}{\partial \ln m_{\mathrm{d}}} + \frac{\partial \ln (R_{\mathrm{RL}}/a)}{\partial \ln q}\frac{\partial \ln q}{\partial \ln m_{\mathrm{d}}}
\end{equation}

\noindent where $R_{\mathrm{RL}}$ is the Roche lobe radius. Through $\zeta_{\mathrm{RL}}$'s dependence on the mass ratio, the stability of mass transfer from the donor depends on the conservation factor $\beta$, the fraction of mass lost by the donor which actually lands on the accretor. During case B mass transfer a long slow nuclear timescale phase is preceeded by a rapid mass loss phase on the thermal timescale \citep{vandenHeuvel1994}. For our study we will take $\beta = 0.3$ assuming that significant mass is lost from the system during this first stage of mass loss. The only non-compact objects we see to accrete at anything comparable to the thermal timescale are Herbig-Haro objects, in which we see the formation of bipolar outflows, and so we assume this is the mechanism by which matter is lost from the system here, as the accretor remains always well within its Roche lobe throughout stable mass transfer for the binaries considered here \cite[see][]{Woods2011arxiv}. Mass is then lost carrying the specific angular momentum of the accretor. We find that this generally allows for stable MT for $0.83 \lesssim q \lesssim 0.92$ and primary masses 1.0 -- 1.3$M_{\odot}$.

\subsection{Double White Dwarf Evolution}  

We can then evolve a set of 1.2+1.1$M_{\odot}$ binaries with varying initial periods through an initial phase of stable non-conservative case B mass transfer using a version of Eggleton's detailed stellar-evolution code {\bf ev} (also referred to as STARS, \cite{Egg71, Egg72, Pols95}). This leaves us with a set of systems in which the initial primary is stripped of its envelope, leaving a proto-helium WD remnant, an expanded orbit, and a companion which will necessarily enter into a CE phase upon its reaching the giant branch and RLOF, due to the dramatically upset mass ratio. We can then determine the result of a CE event by computing the Roche radius and the appropriate core mass for the companion to have reached that size, as well as using eq. 1 with $\alpha _{\mathrm{CE}} = 0.5$. Doing so we find we are able to produce a string of models well in line with the observed distribution of DWDs whose older companion is the less massive, and in which neither component has reached the helium flash \citep[see fig. 2, as well as][]{Woods2011arxiv}.

\begin{center}
\begin{figure}
\includegraphics[height=0.25\textheight]{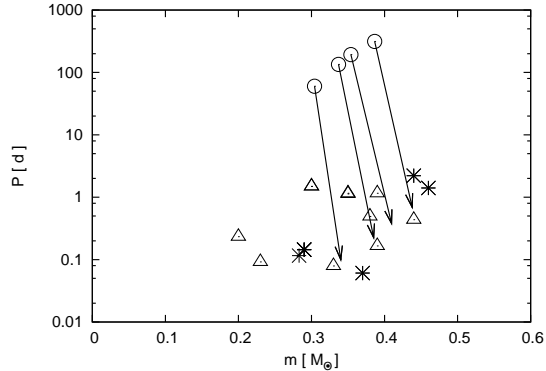}
\caption{Period and component mass for four example binaries evolved from 1.2+1.1$M_{\odot}$ systems with varying period, plotted just prior to the onset of a CE (period and first-born WD mass, circles) and after the CE phase (period and mass of younger WD, end of arrows). Known periods and masses of younger WDs are shown for single-lined (triangles) and double-lined systems (asterisks) as summarized in \cite{WDlist} and \cite{newDWD}.}
\end{figure}
\end{center}

While promising, stable non-conservative MT as followed by a CE can at least explain a subset of the observed population of low-mass DWDs. The extreme effectiveness with which a CE phase removes angular momentum from the system means evolution through two such instances appears unlikely to allow for any binary to avoid merger without very careful fine tuning. Future studies will need to address other potential channels for DWD formation, in order to account for the entire population. 

\acknowledgements We thank P.P. Eggleton and E. Glebbeek for making
their binary-evolution code available to us, as well as Gijs
Nelemans and Craig Heinke for helpful discussion. NI acknowledges support from NSERC and Canada Research
Chairs Program. MvdS acknowledges support from a
CITA National Fellowship to the University of Alberta,
and support from the Dutch Foundation for Fundamental Research on Matter

\end{document}